# Automated Delineation of Hospital Service Areas and Hospital Referral Regions by Modularity Optimization[1]

Yujie Hu[1], Fahui Wang[2*], Imam M Xierali[2,3]
[1]Kinder Institute for Urban Research, Rice University, Houston, TX
[2]Department of Geography and Anthropology, Louisiana State University, Baton Rouge, LA 70803
[3]Association of American Medical Colleges, Washington, DC

**Abstract**

**Objective.** To develop an automated, data-driven, and scale-flexible method to delineate HSAs and HRRs that are up-to-date, representative of all patients, and have the optimal localization of hospital visits.

**Data Sources.** The 2011 State Inpatient Database (SID) in Florida from the Healthcare Cost and Utilization Project (HCUP).

**Study Design.** A network optimization method was used to redefine HSAs and HRRs by maximizing patient-to-hospital flows within each HSA/HRR while minimizing flows between them. We first constructed as many HSAs/HRRs as existing Dartmouth units in Florida, and then compared the two by various metrics. Next, we sought to derive the optimal numbers and configurations of HSAs/HRRs that best reflect the modularity of hospitalization patterns in Florida.

**Principal Findings.** The HSAs/HRRs by our method are favored over the Dartmouth units in balance of region size and market structure, shape, and most importantly, local hospitalization.

**Conclusions.** The new method is automated, scale-flexible, and effective in capturing the natural structure of healthcare system. It has great potential for applications in delineating other healthcare service areas or in larger geographic regions.

**Key Words.** Hospital Service Area (HSA), Hospital Referral Region (HRR), HCUP, Community detection method, Dartmouth method

## INTRODUCTION

A reliable geographic unit is critical to researchers, practitioners, and policy makers involved in healthcare economics and policy, health services and population health research (Kilaru et al. 2015). Such analysis units can be defined by health care markets (e.g., Hospital Service Areas (HSAs) and Hospital Referral Regions (HRRs) provided

---

by Dartmouth Atlas of Health Care, http://www.dartmouthatlas.org/), political boundaries (e.g., county, state), administrative areas (e.g., township), or census units (022e.g., census tracts, metropolitan statistical areas). However, the administrative, census, or political units may be inadequate in healthcare research because they are (1) not intended for healthcare studies, and (2) prone to the modifiable areal unit problem (MAUP) as a result of the multiple levels of aggregation provided, although ancillary data sources are usually available at these units. Instead, the Dartmouth HSAs/HRRs are designed to capture local hospitalization patterns and thus are chosen by many studies as the adequate analysis units for healthcare research such as to examine the geographic variation of the healthcare system (Klauss et al. 2005; AHA 2009; Zuckerman et al. 2010; MedPAC 2011, 2014; Zhang et al. 2012; Newhouse et al. 2013; Kilaru et al. 2015).

The Dartmouth HSAs/HRRs were defined through a three-step process. The first step assigns all acute care hospitals to the town or city where they are located. Based on the 1992-1993 Medicare hospitalization records, the second step uses a plurality rule that assigns each ZIP code to the town or city containing the hospitals most often used by local residents; and the set of ZIP codes assigned to a town or city becomes a preliminary HSA. The third step examines the geographic contiguity of all ZIP codes in a HSA, and assigns, if any, enclave ZIP code(s) to its adjacent HSAs. In a similar fashion, the larger HRRs were subsequently constructed from HSAs based on cardiovascular surgery and neurosurgery referral patterns using the same data (Cooper 1996).

However, the Dartmouth HSAs and HRRs are not free of concerns and need to be improved and updated (Guagliardo et al. 2004; Jia et al. 2015; Kilaru et al. 2015). The first concern is that they were defined about two decades ago; and the hospitalization patterns of residents may have altered significantly over time as a result of the changes of hospitals (e.g., hospitals close, merge, or open), population distribution, infrastructure (e.g., new roads), political regulations, and insurance policies. For example, three of the 114 Dartmouth HSA units in Florida no longer have any hospital in it. These units no longer reflect current hospital utilization patterns.

Secondly, the representativeness of Dartmouth units in terms of patient groups is questioned. As mentioned, the Dartmouth HSAs/HRRs were based on the Medicare patients only and hence were not representative of other patient groups. Guagliardo et al. (2004) assessed the fit of the Dartmouth HSAs for pediatric patients in California and found that they were not appropriate for all age groups and service types. Jia et al. (2015) compared the HSAs derived from the Medicare records with the HSAs based on all-payer data in Florida in 2011, and found significantly different HSA configurations. Kilaru et al. (2015) discovered that Dartmouth units would vary significantly in characterizing the inpatient hospital care patterns by different patient attributes (e.g., insurance type and the utilization of emergency department services) as well as HSA characteristics (e.g., number of hospitals and urban or rural location).

The final concern refers to the scientific soundness of Dartmouth approach for HSA/HRR delineation. The Dartmouth method and other similar approaches (e.g., Klauss et al. 2005; Jia et al. 2015) involve uncertainty or arbitrary choices (e.g., assigning a HSA or enclave to one of its adjacent HSAs to ensure geographic contiguity), and the processes are not completely automated. Additionally, without an explicit objective function in the delineation process, one cannot guarantee that the derived HSAs are optimal and robust in terms of their structures and connections. Given the very goal of defining HSAs for promoting hospital localization, Dartmouth HSAs fell short in ensuring that the patient-to-hospital flows are maximally localized within the HSAs.

The main goal of this study is to develop a completely automated, data-driven, and scale-flexible method to re-delineate HSAs and HRRs that are up-to-date, representative of all patients, and have the optimal hospital localization patterns. The proposed method is based on an optimization technique often used in social network studies to define communities (regions) by repeatedly merging sub-regions (e.g., ZIP code areas) that maximize intra-regional flows (e.g., patient-to-hospital flows) and minimize inter-regional flows (Zhao et al. 2011). In essence, it is a data-driven approach that more accurately detects patient communities hidden in the data, and can be automated in a program for easy implementation. In addition, this method is scale-flexible and can produce a number of HSAs (or HRRs) defined by the user, an important feature that can help policy analysts to examine policy effectiveness at multiple scales. The method is illustrated in a case study in Florida. Various indices are used to demonstrate the benefits of the method.

## Methods
### Study Area and Data Sources
The study area is the state of Florida. Florida is only contiguously connected with two states—Alabama and Georgia—to the north, and bordered by Gulf of Mexico to the west, Atlantic Ocean to the east, and Straits of Florida to the south. This unique geography makes Florida an ideal study area as only a very small population seeks hospital care outside of the state (Jia et al. 2015).

The major data sources used in this research is the State Inpatient Databases (SID) of Florida 2011 from the Healthcare Cost and Utilization Project (HCUP) sponsored by the Agency for Healthcare Research and Quality (AHRQ 2011). The SID includes individual inpatient discharge records in terms of all patients, regardless of payer, from community hospitals in Florida 2011; and contains variables such as principal and secondary diagnoses and procedures, payment source (e.g., Medicare,

Medicaid and private insurance), total charges, patient ZIP codes, and a unique hospital identifier (http://www.hcup-us.ahrq.gov/sidoverview.jsp). We linked the data with the 2013 American Hospital Association (AHA) survey based on the unique hospital identifier, and appended hospital information such as hospital ZIP codes, bed size and hospital type. The Florida 2011 SID consists of 2,656,249 inpatient discharge records from 281 hospitals, including hospital transfers. For patients admitted to multiple hospitals, each admission is regarded as a separate record in this analysis. We excluded 17,178 records associated with 13 hospitals that cannot be identified, 22,733 records without residence ZIP codes, 174,004 observations from hospitals that are not defined as general medical and surgical hospitals by the AHA survey (e.g., pediatric, psychiatric, long-term acute care, rehabilitation, women's and other specialty centers), and 96,302 admissions from out-of-state patients. After these preprocessing, there remained 2,346,032 all-payer records (88.32% of the original records) associated with 202 hospitals (72% of all hospitals) and 983 ZIP codes (100% of all ZIP codes) in Florida in 2011. ZIP code areas were represented by their population-weighted centroids (calibrated from population data at the census block level) for improved spatial accuracy especially for rural ZIP code areas of large area size (Wang, 2015:78). See supplementary Figure S1 for more details about our study population.

**The Community Detection Method**

This research applies the Louvain community detection method (Blondel et al. 2008) to delineate HSAs and HRRs. Often used in social network studies, it is an optimization process of partitioning nodes in a network into natural groups (communities) of nodes (Newman 2004a, 2010). Basically, this approach is an agglomerative hierarchical clustering (i.e., bottom-up) approach. In the beginning, the algorithm treats every node as a group (community), and then successively combines communities together to form larger communities, choosing at each step the best agglomeration with respect to the previous community configuration, until all nodes in the network are grouped into one large community or no improvements in the community configuration are observed.

To examine the quality of each agglomeration (i.e., configuration of the resulted community), Newman and Girvan (2004) designed the so-called modularity, which is now the most acceptable quality measure. Similar to the principle of statistical significance test, it compares the total number of edges (or total weights of edges for a weighted network) fallen within all communities in a given network to that in a null model (i.e., random network), and a good division sees more within-community edges relative to the number of such edges expected by chance (Newman 2004a). Mathematically, the modularity Q of a network (weighted) is formulated as (Newman 2004b)

$$Q = \frac{1}{2m} \sum_{ij} \left( A_{ij} - \frac{k_i k_j}{2m} \right) \delta(c_i, c_j) \tag{1}$$

where $A_{ij}$ represents the edge weight between nodes i and j, $m = \frac{1}{2}\Sigma_{ij}A_{ij}$ is the sum of weights of all edges in the network, $k_i = \Sigma_j A_{ij}$ is the sum of weights of edges linked to node i (also termed the degree of node i), $c_i$ is the community to which node i is assigned, and $\delta(x, y)$ equals 1 when $x = y$, and 0 otherwise. In essence, this equation calculates the difference of total within-community edge weights between a real flow network and an expected flow network, and a higher modularity score indicates a better division (or more steady community structure). A modularity score of 0 indicates that the fraction of total edge weights is no better than that expected at chance, and thus no communities exist in the network; and a score of 1 represents the most robust community structure. In practice, the modularity Q ranges between 0.3 and 0.7 for most real world networks (Newman and Girvan 2004; Newman 2006). As a result, community detection is eventually a modularity optimization process searching for the agglomeration that produces communities with the maximal modularity score, in other words, maximal intra-community flows and minimal inter-community flows (Zhao et al. 2011).

Here, we illustrate the implementation of Louvain community detection method in delineating HSAs in Florida based on its 2011 SID. First, the extracted all-payer patient-to-hospital flows from the SID are used to construct a network with (1) ZIP code (patient residential ZIP codes and hospital ZIP codes) centroids as its nodes, (2) flows between a patient ZIP code and a hospital ZIP code as its edges, and (3) the flow volume as the edge weights. This network is represented by two ASCII files—one for nodes and the other for edges between them—and then fed into the community detection algorithm. Figure 1 illustrates the preceding flow network constructed from Florida 2011 SID. More flows are observed in central Florida and along the western and eastern coastal lines where major metropolitan areas are located. One may also observe the distance decay effect that the flow volumes decrease as the distances between patients and hospitals increase.

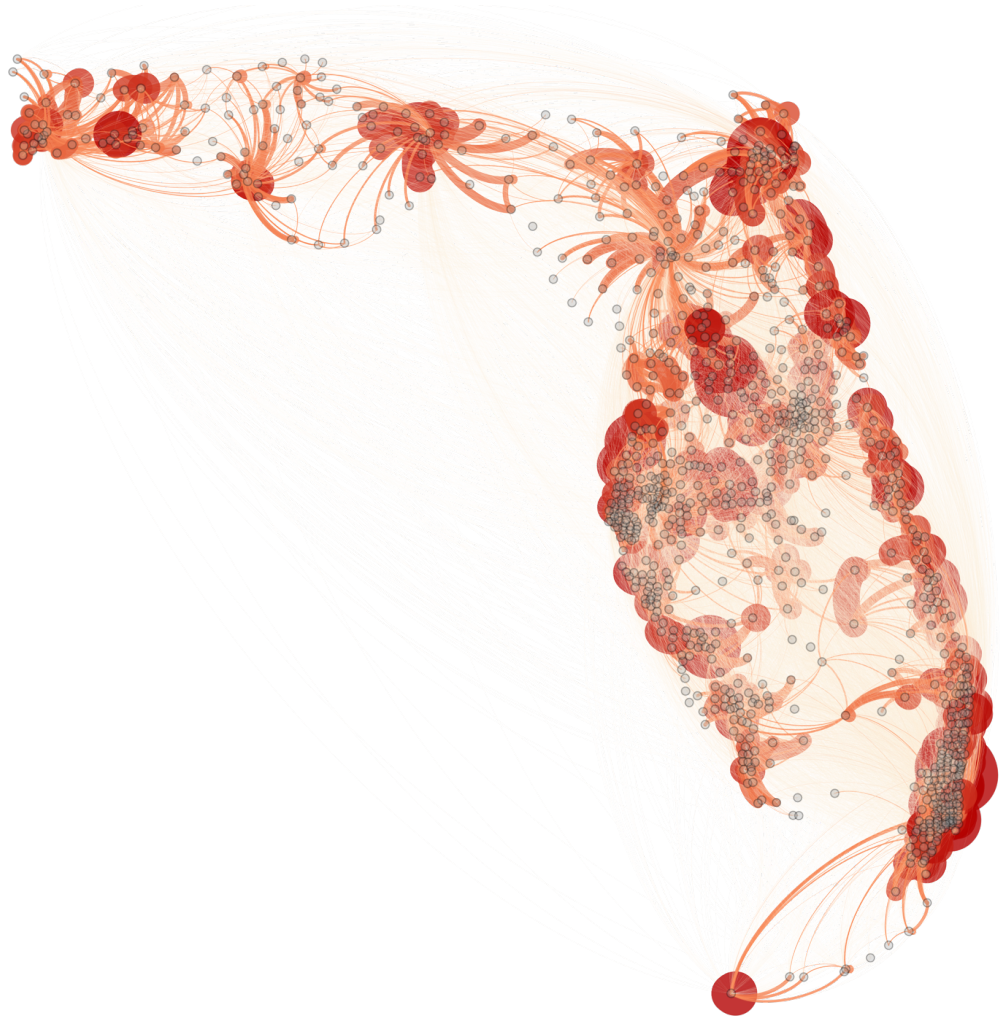

Figure 1. All payer-based patient-to-hospital flow network in Florida 2011
[line width is proportional to the flow volume on it]

Specifically, this algorithm generally iterates based on a two-phase process. The first phase treats each ZIP code centroid (for simplicity, we use node hereafter) in the network as a unique community (e.g., HSA), then for each node i removes it from its original community and puts it into one of its neighboring communities. A local modularity gain is calculated to help identify the destination community for node i that gives the biggest increase in modularity. Equation 2 presents the local modularity gain by adding node i into one of its neighboring communities $C_j$ (Blondel et al. 2008):

$$\Delta Q = \left[\frac{\Sigma_{in}+k_{i,in}}{2m} - \left(\frac{\Sigma_{total}+k_i}{2m}\right)^2\right] - \left[\frac{\Sigma_{in}}{2m} - \left(\frac{\Sigma_{total}}{2m}\right)^2 - \left(\frac{k_i}{2m}\right)^2\right] \quad (2)$$

where $\Sigma_{in}$ is the sum of weights of all edges inside $C_j$, $\Sigma_{total}$ is the sum of weights of all edges that have one of their ends in $C_j$, $k_i$ is the sum of weights of edges linking to node i, and $k_{i,in}$ represents the sum of weights of edges from node i to all nodes in $C_j$. This process is repeated for all nodes until no modularity gain can be obtained. However, the community configuration (i.e., overall modularity) returned by the first phase here is merely a local optimum in this hierarchical approach. The second phase then starts to detect the global optimal result. Specifically, it generates a new network where its nodes are the communities identified in the first phase, and the weights of the edges between the new nodes are the sum of the weight of edges between nodes in two original communities discovered in the first phase. The first phase is then reapplied afterwards. Repeat this two-phase process until there is only one large community left or no positive modularity (overall) change can be experienced. Figure 2 shows the workflow of this method.

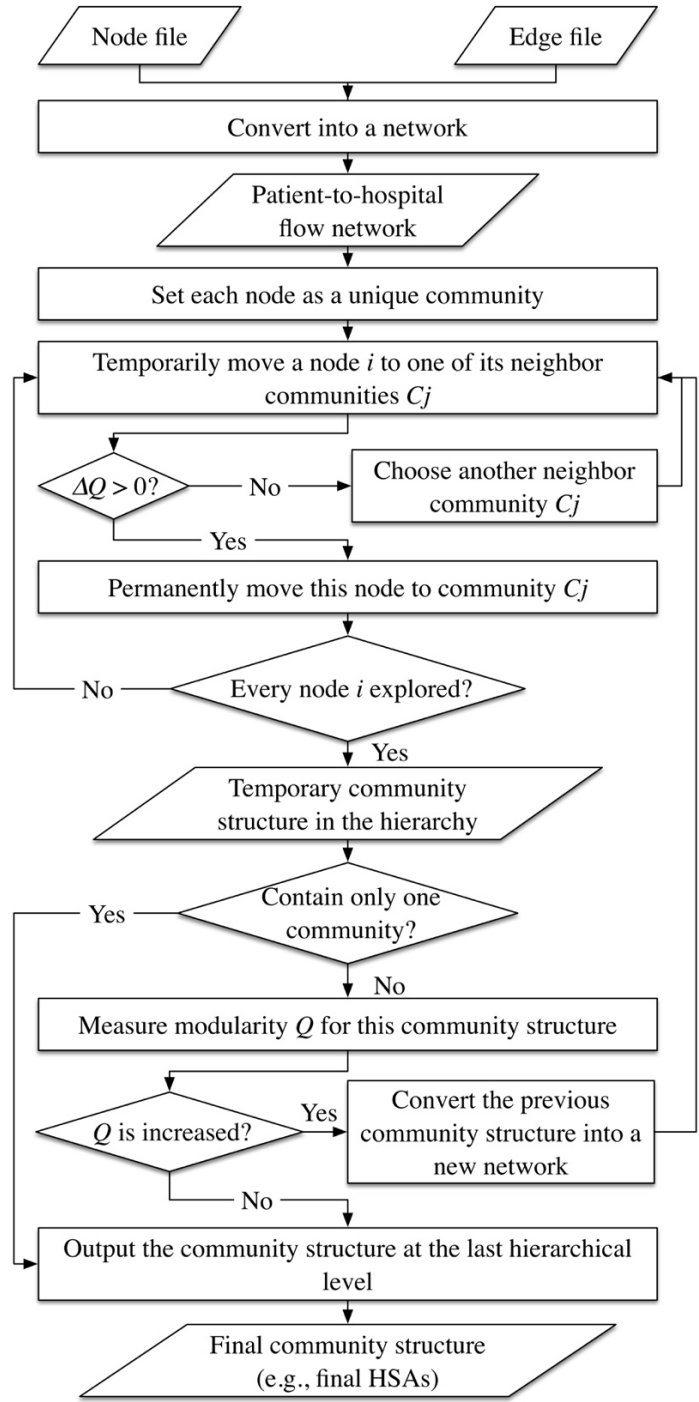

Figure 2. Workflow of the proposed method

From the extracted 2,346,032 records, we further pulled out those records associated with cardiovascular surgery and neurosurgery; and then applied the above workflow again to delineate HRRs in Florida.

In addition to this Louvain method, there are also other methods for community detection in the literature (see reviews in Schaeffer 2007; Fortunato 2010; and Newman 2010). But this approach has one significant feature over others that it is scale-flexible. Specifically, it records every hierarchy of community structures and thus facilitates the investigation of network communities at multiple scales (Fortunato 2010; Ratti et al. 2010). That being said, one may trace back at all hierarchical levels to search for the desired community structure. This scale-flexible nature can be useful to researchers and policymakers to examine the effectiveness of planning and policies targeted at multiple scales or the impacts of MAUP.

**Evaluation Metrics**

We selected nine metrics to characterize HSAs/HRRs and evaluate the performance of the proposed method. These metrics include localization index, market share index, and net patient flow that are commonly used to characterize healthcare service regions (Kilaru et al. 2015), as well as geometric compactness, region size balance, and hospital market structure which are common measures to examine regionalization products from a geographic perspective (Guo 2008). Being the most widely-used indicator that reveals local hospitalization patterns, localization index (LI) describes the proportion of patients that are treated in the same HSA/HRR as where they live, and is designed to capture the propensity of patients visiting local hospitals (Guagliardo et al. 2004). A higher LI demonstrates more accurate or representative delineation of HSA or HRR boundaries. Market share index (MSI) is the proportion of HSA- or HRR-patients who do not live in the regions (Kilaru et al. 2015). It represents the tendency of hospitals in a HSA (or HRR) to absorb out-of-area residents; and hence a lower value generally means better delineation. Net patient flow (NPF) is defined as the ratio of incoming patients to outgoing patients, i.e., the non-HSA (or HRR) residents treated inside the region vs. HSA (or HRR) residents treated outside the region (Klauss et al. 2005). A value $> 1$ indicates the tendency of more patients travelling inside the area to seek hospital care, while a value $< 1$ implies the tendency of more patients leaving the area for hospital care.

Different from these metrics targeted on the healthcare-related characteristics, the following focuses on the geographic structure of HSAs/HRRs. For example, geographic compactness, which is commonly used in evaluating redistricting plans (Siegel 1996), characterizes the regularity of a region's shape based on the perimeter-area corrected ratio or PAC ($= P/(3.54*sqrt(A))$). A geographically compact region (i.e., with lower PAC value) indicates that it is consolidated rather than spread out, to put it another way, packed around its central point (Shirabe and Tomlin 2002); and is proven to benefit systems planning (Mu and Wang 2008).

Balance in region sizes, e.g., relatively even numbers of subregions (here, ZIP codes), hospitals, patients and population in general across HSAs (or HRRs), leads to regions that are more comparable. This is an important property desirable for any regionalization method (Guo 2008) so that constructed regions are more representative of fair sampling in statistical analysis (Wang et al. 2012). It avoids the necessity of a mitigation measure adopted in some studies to assign various weights to samples that differ drastically in population (Wang, 2015:122-126).

Herfindahl Index (HHI), an economic concept measuring the amount of competition among firms in a local market, is adopted here to characterize the hospital market structure, specifically, the competitive environment in a HSA (or HRR) in terms of share of admissions to local hospitals. Investigations into HHI may help researchers and policymakers better understand hospital behaviors and answer such questions as "are there higher costs or more specialized services, on average, when hospitals are located in more competitive hospital market areas? (Garnick et al. 1987)" HHI is the sum of the squared market share of each hospital in the market (i.e., HSA or HRR) multiplied by 10,000, where the market share of a hospital is its share of inpatient admissions in the same HSA (or HRR). For example, a HSA with only one hospital would have a HHI of 10,000. On the contrary, a HSA with a great number of hospitals that have relatively even shares of inpatient admissions would have a HHI near 0. As is standard, a hospital market is considered highly concentrated (or monopolistic) if the HHI is larger than 2,500, moderately concentrated if between 1,500 and 2,500, unconcentrated if between 100 and 1,500, and highly competitive if below 100 (Cutler and Morton 2013).

### RESULTS AND COMPARISON TO THE DARTMOUTH UNITS

An experiment is designed in order to evaluate the performance of our method with comparison to the Dartmouth method. First, we used the new method to construct as many HSAs/HRRs as the Dartmouth units in Florida, hereafter referred to as Dartmouth-comparable units for simplicity—thanks to the scale-flexible feature of this method. We then applied the aforementioned metrics to characterize and compare the two units. Given that the Dartmouth-comparable units may not have the global optimal modularity, the next section derived the HSAs/HRRs with global optimal modularity value and applied those metrics to detect the "optimal" (or "natural") configuration of a hierarchical hospitalization system in Florida.

In Florida, there are 114 HSAs and 18 HRRs delineated by Dartmouth. Therefore, we constructed 114 Dartmouth-comparable HSAs and 18 Dartmouth-comparable HRRs using the all-payer based patient-to-hospital flow network (see

Figure 1) and the subset of only major cardiovascular surgical or neurosurgery flow network, respectively. Given the modularity score range 0.3-0.7 in the corresponding real world networks, the modularity scores for the Dartmouth-comparable units (0.63 for HSAs and 0.8 for HRRs) were notably high and indicated the significant non-random region structures of newly-derived units. The following evaluated the effectiveness of these units in comparison to Dartmouth ones in the nine indices.

Localization index (LI) is the proportion of patients that are treated in the same HSA/HRR as where they live, and a higher value represents more favorable delineation. As shown in Figure 3a, the mean LI in Dartmouth-comparable HSAs is slightly higher than that in the Dartmouth HSAs (0.54 vs. 0.53), though not statistically significant. For HRRs (see Figure 4a), Dartmouth-comparable HRRs return a higher mean LI than Dartmouth HRRs (0.9 vs. 0.87) and the difference is significant at the 0.01 level. In addition, a significantly smaller range of LI is observed in both Dartmouth-comparable HSAs and HRRs. That is to say, our modularity optimization approach shows an advantage in improving hospitalization LI, and such advantage is more evident in HRRs than HSAs.

Market share index (MSI) reflects the proportion of HSA- or HRR-patients who do not live in the regions and therefore a lower value indicates better delineation. As illustrated in Figures 3b and 4b, we find lower mean MSI (significant at 0.01 level) in Dartmouth-comparable units than Dartmouth ones (0.33 vs. 0.37 for HSAs, and 0.09 vs. 0.11 for HRRs). Likewise, it also indicates the better performance of our method in delineating hospital service regions with improved local healthcare patterns.

Net patient flow (NPF) is the ratio of incoming patients to outgoing patients, and a value closer to 1 indicates more balanced healthcare service structure, i.e., better localization pattern in the region. Both Figures 3c and 4c show lower mean NPF value in Dartmouth-comparable units than Dartmouth's, for example, 1.15 vs. 1.24 for HSAs and 1.09 vs. 1.17 for HRRs, and the differences are significant at 0.01. Again, the less deviation of NPFs from 1 as well as the narrower range of NPFs in Dartmouth-comparable units demonstrate improved local healthcare patterns.

Compactness here is measured in PAC as explained previously. As shown in Figures 3d and 4d, the Dartmouth-comparable HSAs and HRRs have overall lower mean PAC values than Dartmouth ones (2.18 vs. 2.4 for HSAs, and 3.1 vs. 3.44 for HRRs, both significant at 0.01), indicating a more regularly-shaped and consolidated region of the newly constructed HSAs and HRRs. The wider variability in compactness of Dartmouth HSAs is especially noticeable.

Balance in region size is measured in four aspects such as sub-region count, numbers of hospitals and inpatients, and population size in this study. In practice, relatively even region sizes are often desired by planners (Guo 2008). Figures 3e and 4e show more balanced sub-region counts (here, referred to simply as region sizes) in

Dartmouth-comparable units than in Dartmouth ones. The region sizes for Dartmouth-comparable units have narrower ranges and lower standard deviations than those for Dartmouth units. As we generated the same number of HSAs (and HRRs) as Dartmouth units, the mean region size between two HSAs (and HRRs) is identical. Also as shown in Figures 3f-3h and 4f-4h, the number of hospitals in HSAs (and HRRs) varies in a significantly narrower range with a lower standard deviation in Dartmouth-comparable HSAs (and HRRs) than Dartmouth units. Similar findings are observed for the number of inpatients and population size in HSAs (and HRRs). That is to say, our Dartmouth-comparable units are more balanced in their sizes, in hospital count as well as the inpatients and population served, and therefore are more comparable across HSAs or across HRRs.

Herfindahl Index (HHI) measures the competitive environment in a HSA (or HRR) in terms of share of admissions to local hospitals, and a smaller value indicates more competition (i.e., more balanced share of inpatient admissions) among hospitals in the same region. A highly concentrated or monopolistic market structure is observed in Dartmouth HSAs (mean HHI = 3510); however, only moderately concentrated market structure in Dartmouth-comparable HSAs (mean HHI = 2275). At the HRR level, both units have unconcentrated markets, but Dartmouth-comparable HRRs (mean HHI = 162) are more competitive in structure than Dartmouth HRRs (mean HHI = 255). See Figures 3i and 4i for more details.

In summary, the newly-derived HSAs and HRRs are more favorable than the Dartmouth units in all nine indices used: localization of hospital visits (in LI, MSI and NPF), compactness in shape, balanced region sizes (in four measures) and market structure.

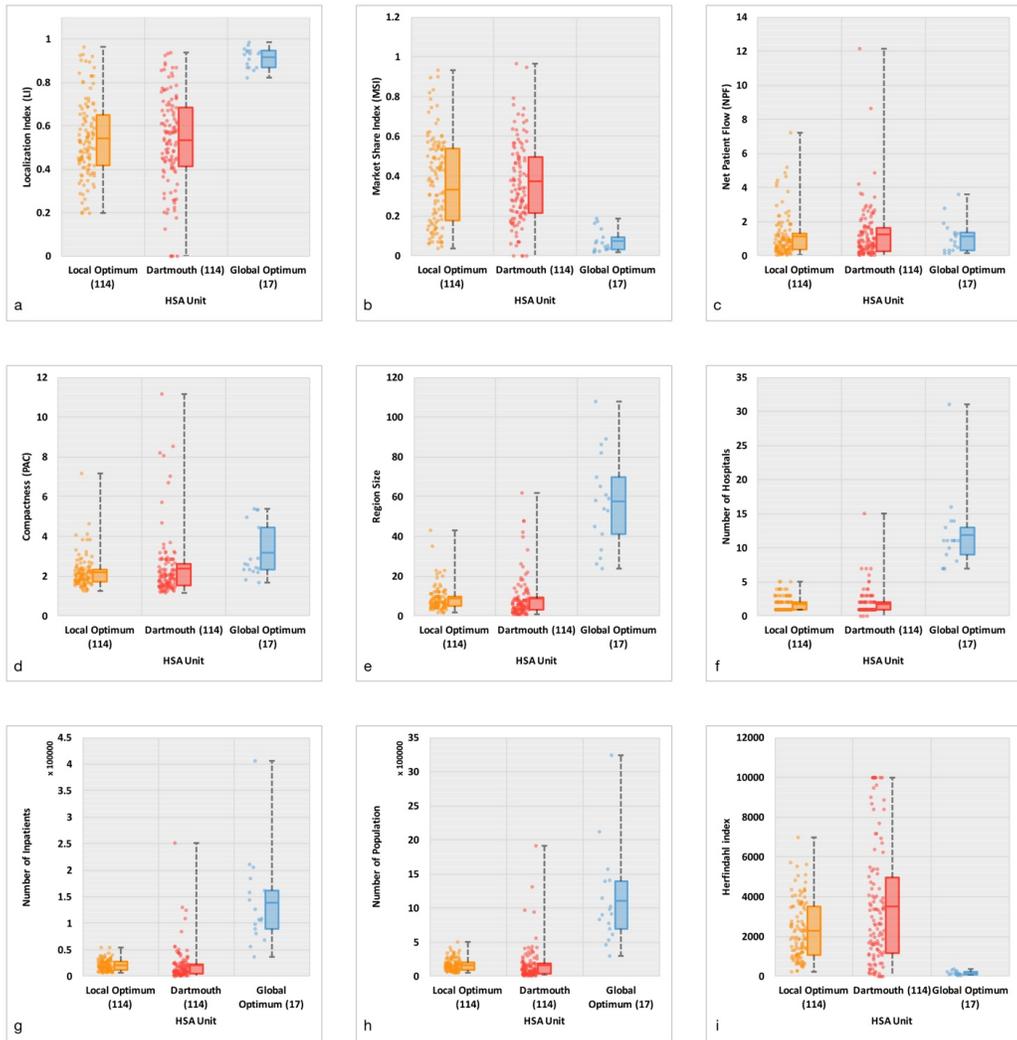

Figure 3. Basic statistics of nine criteria for evaluating HSAs
[The line in the midst of a box represents mean not median value]

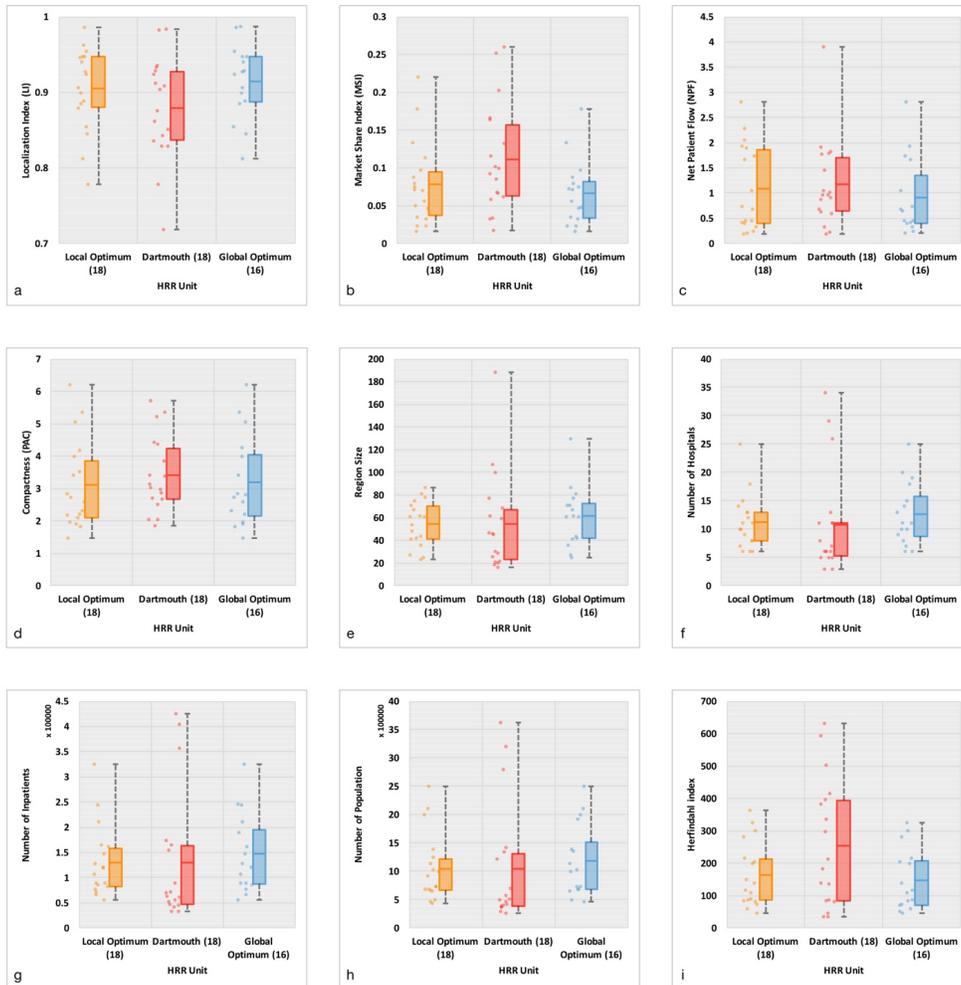

Figure 4. Basic statistics of nine criteria for evaluating HRRs
[The line in the midst of a box represents mean not median value]

## Exploratory analysis of optimal configuration of hospital service areas

As mentioned previously, the Dartmouth-comparable units may not represent the best feasible configuration of hospital service market in Florida. In fact, the 114 Dartmouth-comparable HSAs and 18 HRRs are geographic units with only local optimal modularity scores. In other words, is there an optimal number of HSAs (or HRRs) that best capture the "natural" structure of healthcare system?

Based on the patient-to-hospital flow pattern for all patients, our method yielded 17 HSAs with the global optimal modularity score of 0.85 (see supplementary Figure S2), referred to as "global-optimal HSAs" hereafter. For the cardiac and neuro surgery patient flow pattern that were used for delineation of HRRs, our method yielded 16 HRRs with the global optimal modularity score of 0.83 (the graph for modularity scores vs. HRR size is not shown here), referred to as "global-optimal HRRs" hereafter. Note that the number of global-optimal HRRs derived from the specialized care patient discharge pattern was just one fewer than the number of global-optimal HSAs based on all patients, and their configurations (17 HSAs vs. 16 HRRs) were largely consistent (see Figures 5a and 5b). Both numbers were also very close to the 18 Dartmouth HRRs. To our knowledge, such an analysis in the context of healthcare market is the first of the kind, and thus both the results and related discussion are exploratory in nature.

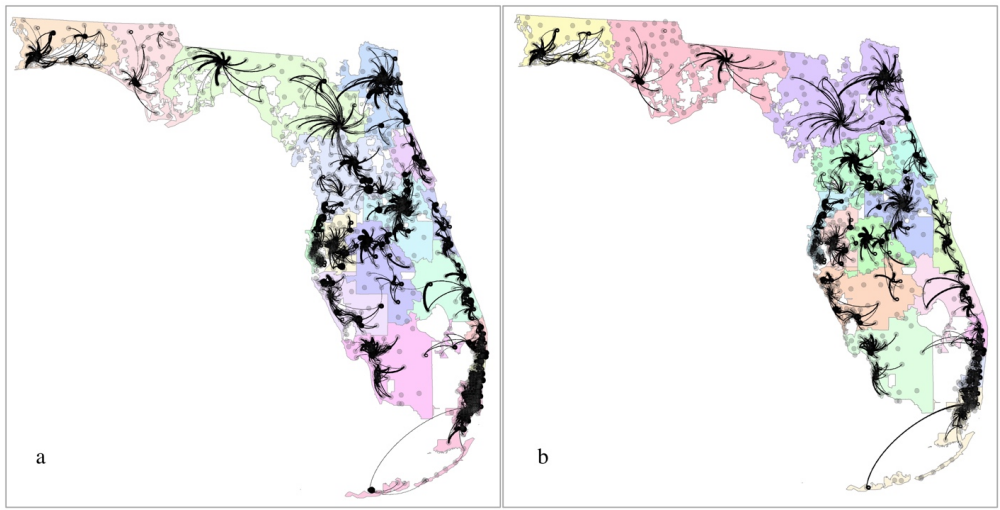

Figure 5. Major discharge flows and global optimal units in Florida: a. HSAs with ≥ 200 inpatient discharge and b. HRRs with ≥ 50 inpatient discharge
[White space is water areas]

The most remarkable improvement is observed in healthcare localization pattern. The mean LI increased from 0.54 in the 114 Dartmouth-comparable HSAs (LI=0.53 for the 114 traditional Dartmouth HSAs) to 0.92 in the 17 global-optimal HSAs. Granted that much of the improvement in LI was attributable to the aggregation to much fewer HSAs, the significant increase of modularity score from a local optimal modularity score 0.63 to a global optimal 0.85 signaled the need of further consolidation of HSAs in order to capture the contemporary hospitalization pattern that is increasingly interwoven and integrated. That is to say, the number of HSAs defined by the traditional Dartmouth method might be too high (i.e., the size of Dartmouth HSAs might be too small) for today's healthcare market. A similar increasing trend, understandably not as drastic, is detected in HRRs: the mean LI increased from 0.87 in the 18 Dartmouth HRRs to 0.90 in the 18 Dartmouth-comparable HSAs, and then to 0.92 in the 16 global-optimal HRRs. Clearly, both the aggregation to slightly fewer HRRs and the optimal configuration of internal patient flows contributed to the improved LI. As for the second related index MSI, we found a significant decline in the mean MSI from 0.33 in Dartmouth-comparable HSAs to 0.07 in global optimal HSAs. In other words, if the 17 global-optimal HSAs were adopted in Florida, hospitalization was highly localized with only 7 percent of patients visiting hospitals outside their assigned HSAs. Similarly, there were an overall 9 percent patients visiting outside the 18 Dartmouth-comparable HRRs and also 7 percent outside in the 16 global-optimal HRRs. Once again, the convergence of global-optimal HSAs and global-optimal HRRs (though one based on all patients and another on "specialized" patients) is noticeable and has important implication. The same message can be read from the NPF results: the mean value was 1.12, 1.15 and 1.24 in the global-optimal, Dartmouth-comparable and Dartmouth HSAs, respectively; and its mean value was 0.94, 1.09 and 1.17 in the global-optimal, Dartmouth-comparable and Dartmouth HRRs, respectively.

Other indices such as balance in region size, shape compactness and market structure are strongly dependent on scale. For example, according to the HHI, the global-optimal HSAs had a significantly more equitable share of patient admissions among local hospitals (mean HHI = 150), compared to a moderately monopolistic market structure in Dartmouth-comparable HSAs (mean HHI = 2275). This is mainly caused by the major difference in number of hospitals enclosed in the two systems (i.e., smaller and more numerous HSAs in a Dartmouth-comparable system are more likely to be dominated by one or a few hospitals). It is not very meaningful to evaluate the differences between 114 Dartmouth-comparable HSAs and 17 global-optimal HSAs in those indices. Interested readers may refer to Figures 3 and 4 (the 3 boxplots correspond to Dartmouth-comparable, Dartmouth and global-optimal units in each graph).

To recap this exploratory analysis, Figures 5a and 5b show the global-optimal HSAs (only discharge flows ≥ 200 inpatient discharge are preserved to highlight the pattern) and HRRs (only flows ≥ 50 "specialized" inpatient discharge are kept). Clearly, both flow maps appear to align well with the delineated geographic units. Specifically, we find more patient-to-hospital flows within a region and negligible flows between regions, and hence a highly localized healthcare service pattern.

CONCLUDING REMARKS

Several issues may be identified in the widely-used Dartmouth HSAs and HRRs as analysis units for healthcare market. They are outdated and unrepresentative of the overall patient population, and the delineation process is not automated and involves making some arbitrary choices. Based on 2011 SID data in Florida, the primary purpose of this research is to adopt a community detection method developed in the complex network science for delineating HSAs and HRRs that are (1) up-to-date, (2) representative of the overall population, and (3) automatically and consistently generated. The derived HSAs/HRRs possess significant advantages over the Dartmouth units. Our units are more balanced in region size and more compact in shape, two important properties in design of any regionalization method. Most importantly, our units enjoy higher local hospitalization ratios, the very foundation built in any approach for hospital service area delineation.

The exploratory analysis of seeking the optimal number of HSAs also yields some interesting findings and sheds light on several important issues in the field. First, the global-optimal HSAs (in terms of modularity in existing overall patient flows) has a number very close to the number of HRRs, either based on the optimized modularity of cardiac and neuro surgery patient flows or from the traditional Dartmouth method. This convergence may suggest the need of revisiting the use of these two services in defining the HRRs. The increasing prevalence of such services (especially cardiac surgery) might have reached a point that they are no longer as specialized or with limited availability as they used to be. A recent study (Jia, 2015) shows similar distance decay behaviors in general inpatients and inpatients of cardiac procedures in Florida hospitals. Secondly, by extension of the previous point, one may consider the larger HRRs aggregated from HSAs as a more appropriate unit for analysis of today's healthcare market that is increasingly consolidated as well as integrated. Thirdly, both points call for more studies to be expanded temporally (e.g., to verify whether the healthcare market has indeed become more geographically interwoven over time) and spatially (e.g., in a larger geographic region so that sufficient samples of patients of specialized cares are available for reliable statistical analysis). For now, the above discussion is suggestive or perhaps speculative, and needs to be taken with caution.

There are some limitations in the research. Here we raise several related to the data source. The SID data included information on hospital transfers and we treated each admission as a separate record. For instance, about 20% of the selected 2,346,032 records in Florida were hospital transfers. Inclusion or separation of transfer patients, readmissions and treat-and-release outpatients may affect the results. Future research should consider this to more fully assess the healthcare system and patient communities. In addition, due to data availability, we only used inpatient flows (SID) but not the outpatient flows data. When available, both data sources may be integrated (or analyzed separately) to assess the whole of healthcare service pattern. The geographic level in the SID data is also inconsistent across datasets (e.g., five states are not aggregated at ZIP level). While our method can be applied in any scale, the inconsistency across states is an issue that needs to be resolved for a study area including multiple states. Furthermore, the selected SID data are only available for roughly 50% of states across the US. This raises a problem for feasibility of national analysis. Like the Dartmouth Atlas Project, one may seek other data sources that are provided nationwide such as the CMS Medicare claims data even though it is limited to a cohort of Medicare patients only. This paper focuses on methodological issues of defining HSAs, and we hope to report future work on using the new units to compare the geographic variation in spending, population, utilization, and other measures in the healthcare system as raised in a recent Institute of Medicine report (Newhouse et al. 2013).

**References**
AHA (American Hospital Association). (2009). Geographic Variation in Health Care Spending: A Closer Look. TrendWatch, 2009, 1-16.
Agency for Healthcare Research and Quality (AHRQ). (2011). Healthcare Cost and Utilization Project (HCUP) State Inpatient Databases (SID) - Florida. Rockville, MD. www.hcup-us.ahrq.gov/sidoverview.jsp.
Blondel, V. D., Guillaume, J. L., Lambiotte, R., and Lefebvre, E. (2008). Fast unfolding of communities in large networks. Journal of Statistical Mechanics: Theory and Experiment, 2008(10), P10008.
Cooper, M. M. (1996). The Dartmouth Atlas of Health Care. Chicago, IL: American Hospital Publishing, pp. 2-26.
Cutler, D. M., and Morton, F. S. (2013). Hospitals, market share, and consolidation. JAMA, 310(18), 1964-1970.
Fortunato, S. (2010). Community detection in graphs. Physics Reports, 486(3), 75-174.
Garnick, D. W., Luft, H. S., Robinson, J. C., and Tetreault, J. (1987). Appropriate measures of hospital market areas. Health services research, 22(1), 69-89.
Guagliardo, M. F., Jablonski, K. A., Joseph, J. G., and Goodman, D. C. (2004). Do pediatric hospitalizations have a unique geography?. BMC health services research, 4(1), 2.

Guo, D. (2008). Regionalization with dynamically constrained agglomerative clustering and partitioning (REDCAP). International Journal of Geographical Information Science, 22(7), 801-823.

Jia, P. (2015). Delineating Hospital Service Areas in Florida Based on Patients' Travel Patterns. PhD Dissertation, Louisiana State University.

Jia, P., Xierali, I. M., and Wang, F. (2015). Evaluating and re-demarcating the Hospital Service Areas in Florida. Applied Geography, 60, 248-253.

Kilaru, A. S., Wiebe, D. J., Karp, D. N., Love, J., Kallan, M. J., and Carr, B. G. (2015). Do hospital service areas and hospital referral regions define discrete health care populations?. Medical care, 53(6), 510-516.

Klauss, G., Staub, L., Widmer, M., and Busato, A. (2005). Hospital service areas–a new tool for health care planning in Switzerland. BMC health services research, 5(1), 33.

MedPAC (Medicare Payment Advisory Commission). (2011). Regional variation in Medicare service use. In Report to the Congress: January 2011. Washington, DC: MedPAC.

MedPAC (Medicare Payment Advisory Commission). (2014). Report to the Congress: Medicare payment policy: March 2014. Washington, DC: MedPAC.

Mu, L., and Wang, F. (2008). A scale-space clustering method: Mitigating the effect of scale in the analysis of zone-based data. Annals of the Association of American Geographers, 98(1), 85-101.

Newhouse, J. P., Garber, A. M., Graham, R. P., McCoy, M. A., Mancher, M., and Kibria, A. (Eds.). (2013). Variation in Health Care Spending: Target Decision Making, Not Geography. National Academies Press.

Newman, M. E. (2004a). Fast algorithm for detecting community structure in networks. Physical review E, 69(6), 066133.

Newman, M. E. (2004b). Analysis of weighted networks. Physical Review E, 70(5), 056131.

Newman, M. E. (2006). Modularity and community structure in networks. Proceedings of the National Academy of Sciences, 103(23), 8577-8582.

Newman, M. E. (2010). Networks: an introduction. Oxford University Press.

Newman, M. E., and Girvan, M. (2004). Finding and evaluating community structure in networks. Physical review E, 69(2), 026113.

Ratti, C., Sobolevsky, S., Calabrese, F., Andris, C., Reades, J., Martino, M., Claxton, R., and Strogatz, S. H. (2010). Redrawing the map of Great Britain from a network of human interactions. PloS one, 5(12), e14248.

Schaeffer, S. E. (2007). Graph clustering. Computer Science Review, 1(1), 27-64.

Shirabe, T., and Tomlin, C. D. (2002). Decomposing integer programming models for spatial allocation (pp. 300-312). Springer Berlin Heidelberg.

Siegel, J. S. (1996). Geographic compactness vs. race/ethnic compactness and other criteria in the delineation of legislative districts. Population Research and Policy Review, 15(2), 147-164.

Wang, F. (2015). Quantitative Methods and Socioeconomic Applications in GIS (2$^{nd}$ ed.) Boca Raton, FL: CRC Press

Wang, F., D. Guo and S. McLafferty. (2012). Constructing geographic areas for cancer data analysis: a case study on late-stage breast cancer risk in Illinois. Applied Geography 35: 1-11.

Zhang, Y., Baik, S. H., Fendrick, A. M., and Baicker, K. (2012). Comparing local and regional variation in health care spending. New England Journal of Medicine, 367(18), 1724-1731.

Zhao, Y., Levina, E., and Zhu, J. (2011). Community extraction for social networks. Proceedings of the National Academy of Sciences, 108(18), 7321-7326.

Zuckerman, S., Waidmann, T., Berenson, R., and Hadley, J. (2010). Clarifying sources of geographic differences in Medicare spending. New England Journal of Medicine, 363(1), 54-62.

**Acknowledgements:** The Economic Development Assistantship (EDA) support from the Graduate School of Louisiana State University to Hu is gratefully acknowledged.